\documentclass[prper,aps,twocolumn,showpacs,groupedaddress,floats,final,superscriptaddress]{revtex4-1}
\usepackage{graphicx}
\usepackage{dcolumn}
\usepackage{bm}
\usepackage{color}
\usepackage{amsmath}
\usepackage{amssymb}
\usepackage{amsfonts}
\RequirePackage[
  hyperindex,colorlinks,bookmarksnumbered,
  plainpages=true,pdfstartview=FitH]{hyperref}
\hypersetup{linkcolor=blue,urlcolor=blue,citecolor=blue}
\usepackage{appendix}
\usepackage{hyperref}

\begin{document}

\title{Understanding Quantum Tunneling using Diffusion Monte Carlo Simulations}

\author{E. M. Inack}
\affiliation{The Abdus Salam International Centre for Theoretical Physics, I-34151 Trieste, Italy}
\affiliation{SISSA - International School for Advanced Studies, I-34136 Trieste, Italy}
\affiliation{INFN, Sezione di Trieste, I-34136 Trieste, Italy} 
\author{G. Giudici}
\affiliation{SISSA - International School for Advanced Studies, I-34136 Trieste, Italy}
\affiliation{INFN, Sezione di Trieste, I-34136 Trieste, Italy} 
\author{T. Parolini}
\affiliation{SISSA - International School for Advanced Studies, I-34136 Trieste, Italy}
\affiliation{INFN, Sezione di Trieste, I-34136 Trieste, Italy} 
\author{G. Santoro}
\affiliation{The Abdus Salam International Centre for Theoretical Physics, I-34151 Trieste, Italy}
\affiliation{SISSA - International School for Advanced Studies, I-34136 Trieste, Italy}
\affiliation{CNR-IOM  Democritos  National  Simulation  Center,  Via  Bonomea  265,  34136  Trieste,
Italy} 
\author{S. Pilati}
\affiliation{Dipartimento di Fisica e Astronomia ``Galileo Galilei", Universit{\`a} di Padova, I-35131 Padova, Italy}

\begin{abstract}
In simple ferromagnetic quantum Ising models characterized by an effective double-well energy landscape the characteristic tunneling time of path-integral Monte Carlo (PIMC) simulations has been shown to scale as the incoherent quantum-tunneling time, i.e., as $1/\Delta^2$, where $\Delta$ is the tunneling gap. Since incoherent quantum tunneling is employed by quantum annealers (QAs) to solve optimization problems, this result suggests there is no quantum advantage in using QAs w.r.t. quantum Monte Carlo (QMC) simulations. A counterexample is the recently introduced shamrock model, where topological obstructions cause an exponential slowdown of the PIMC tunneling dynamics with respect to incoherent quantum tunneling, leaving the door open for potential quantum speedup, even for stoquastic models. 
In this work, we investigate the tunneling time of projective QMC simulations based on the diffusion Monte Carlo (DMC) algorithm without guiding functions, showing that it scales as $1/\Delta$, i.e., even more favorably than the incoherent quantum-tunneling time, both in a simple ferromagnetic system and in the more challenging shamrock model. However a careful comparison between the DMC ground-state energies and the exact solution available for the transverse-field Ising chain points at an exponential scaling of the computational cost required to keep a fixed relative error as the system size increases.
\end{abstract}


\maketitle

Difficult optimization problems are ubiquitous in science and in engineering. 
Relevant examples are protein folding, the traveling salesman problem, and portfolio optimization. 
Such problems can often be formulated as the search of the lowest-energy spin configuration in an Ising glass~\cite{lucas2014ising}, a task that has been proven to be $\mathrm{NP}$-hard  in the case of non-planar graphs~\cite{Barahona}.
While exact classical algorithms are believed to require computational times that exponentially grow with the problem size (unless $\mathrm{P} = \mathrm{NP}$), various heuristic methods can often provide quite accurate (but possibly not exact) solutions in a feasible time. Perhaps, the most versatile of such heuristic methods is simulated classical annealing (SCA)~\cite{vecchi}, which exploits thermal fluctuations in a Markov chain Monte Carlo simulation to escape local minima and, hopefully, find the lowest energy state at the end of the annealing process when the temperature has been reduced to zero.
 
Also adiabatic quantum computers, such as the quantum annealers (QAs) built using superconducting flux qubits~\cite{johnson2011quantum,boixo2013experimental,boixo2014evidence} 
--- or, potentially, with Rydberg atoms trapped in arrays of optical tweezers~\cite{glaetzle} --- can be used to solve complex combinatorial optimization problems. 
They implement a quantum annealing process~\cite{Finnila_CPL94,santorotheory,nishimoriising}, 
in which quantum mechanical tunneling through tall barriers is used to escape local minima, and quantum fluctuations are gradually removed by reducing to zero the transverse field of a quantum Ising model.
While in problems with energy landscapes characterized by tall but thin barriers quantum tunneling definitely makes QAs more efficient than classical optimization methods such as SCA~\cite{farhi2002quantum,crosson2016simulated}, 
certain examples are known where the opposite seems to be true~\cite{Battaglia_PRE05,Polkovnikov_PRL15}. 

Giving a definitive answer to the question of which optimization problems can show a definite quantum advantage of some sort~\cite{troyerdefining} is a formidably difficult task, 
since simulating the real time dynamics of QAs using classical computers is feasible only for very small system sizes (up to, say, $\sim 30$ qubits),
which typically tend to be not representative of the real difficulty of a large size problem.
However, quantum annealing can also be implemented on classical computers using quantum Monte Carlo (QMC) algorithms, giving one access to large system sizes.
This approach, which is now often referred to as simulated quantum annealing (SQA)~\cite{santorotheory,santoroPIMC,santorosimplecases,santorodoublewell,troyerheim}, 
represents an alternative heuristic optimization algorithm running on classical computers. It might be competitive with, or even superior to, its classical counterpart SCA. 
The performance of SQA in solving large ensembles of Ising-glass instances has been compared to the one of QAs, finding high correlations between the instances that were easy or hard for the two solvers~\cite{boixo2014evidence,albash2015reexamining}. However, the dynamics of  QMC simulations does not represent the unitary dynamics of a (perfectly isolated) QA; therefore, it is not clear if SQA is a trustworthy probe to predict when QAs may or may not outperform classical optimization algorithms~\cite{santorotheory,troyerheim}.

In a recent study~\cite{isakovtunneling}, which aimed at shedding light on the relation between the dynamics of SQA and the one of QAs, 
it was found that the characteristic timescale of tunneling events occurring during path-integral Monte Carlo (PIMC) simulations increases with the system size as $1/\Delta^2$, where $\Delta$ is the energy gap  between the ground state and the first exited state (see also Ref.~\cite{brady2016quantum}). 
This $1/\Delta^2$ scaling was found to hold in ferromagnetic quantum Ising models~\cite{isakovtunneling}, which are characterized by an effective double-well energy landscape 
(the two symmetric minima are the ground states with opposite magnetizations), and also in one-dimensional and two-dimensional continuous-space double-well models 
relevant for quantum chemistry applications~\cite{mazzolaquantumchemistry}. 
Remarkably, this is the same scaling of the time of \emph{incoherent} quantum tunneling in symmetric double-well models~\cite{weiss1987incoherent}. Furthermore, according to the adiabatic theorem, also the annealing time required in a \emph{coherent} adiabatic quantum computation to avoid diabatic transitions~\cite{farhi2000quantum} to the first exited state must increase as the squared inverse of the smallest instantaneous gap.
The similar scaling of the respective tunneling times --- which was explained using an instanton theory~\cite{jianginstanton} --- suggests that PIMC simulations can efficiently simulate incoherent quantum tunneling. 
This latter phenomenon is supposed to be one of the empowering resources of QAs (although quantum superposition and entanglement might also be crucial ingredients). It allows them to explore different localized states far away in Hamming distance, like those typically emerging in the glassy phases characteristic of Ising glass models at small transverse field. 
On the one hand, this finding suggests that SQA can be used to predict the performance of QAs, providing us with a useful tool to guide the engineering of these devices. On the other hand, it might also imply that SQA has the same potential efficiency in solving complex optimization problems as QAs have, meaning that quantum speedup is unlikely to be achieved (apart for a prefactor), 
at least as long as the Hamiltonian under consideration is stoquastic, i.e., free of any sign problem~\cite{nishimoriNonstoquastic,troyerNonstoquatic}.\\
Later on, Ref.~\cite{aminshamrock} introduced the so-called ``shamrock model'', showing that due to topological obstructions~\cite{hastings2013} --- which originate from frustrated couplings 
in this model --- the PIMC tunneling time increases with the system size exponentially faster than the incoherent quantum tunneling time, giving one hope that QAs can outperform SQA, and so maybe also other heuristic optimization methods running on classical computers.

The performance of SQA in solving optimization problems crucially depends on the specific type of QMC algorithm being used to drive the simulation; 
in particular, for certain difficult optimization problems, a projective QMC method such as the diffusion Monte Carlo (DMC) algorithm has been shown to represent a 
more efficient engine for SQA than the PIMC method~\cite{inack}. 
In fact, SQA optimizations powered by projective QMC methods have proven to be competitive with state-of-the-art classical optimization methods~\cite{jordanDMC}. 
The above mentioned study~\cite{aminshamrock} of the QMC tunneling time for the shamrock model considered only finite-temperature PIMC algorithms. 
This naturally raises the following questions: can projective QMC methods efficiently simulate quantum mechanical tunneling? 
Would they be immune from the (exponential) pathological slowdown which affects the PIMC simulations in the shamrock model? 

The main goal of this paper is to address the above two questions. In order to do so, we implement a projective QMC method for quantum Ising models based on the DMC algorithm in which the stochastic dynamics is defined by the Trotter-decomposed imaginary-time evolution operator.  Then, following Refs.~\cite{isakovtunneling,mazzolaquantumchemistry}, we introduce a protocol to measure the characteristic time of tunneling events occurring in DMC simulations, and we analyze the scaling with the system size of the so-defined tunneling time, both in the ferromagnetic quantum Ising chain and in the shamrock model. 
Furthermore, in order to understand if the DMC algorithm allows one to \emph{efficiently} simulate the behavior of QAs on classical computers, we analyse the computational cost of DMC ground-state simulations. In particular, we study the convergence of the systematic biases in calculations of the ground-state energy, using as a testbed the quantum Ising chain. It should be noted that we focus on the \emph{simple} DMC algorithm, i.e., we do not consider the use of importance sampling techniques~\cite{kalosImportancesampling} based on suitably constructed guiding wave functions.

We find that the DMC tunneling time grows proportionally to the inverse of the gap $1/\Delta$ when the system size increases. This behavior is analogous to what was previously found~\cite{isakovtunneling,mazzolaquantumchemistry} in modified PIMC simulations performed using open-boundary conditions in imaginary time, and it represents a quadratic speedup compared to finite-temperature PIMC simulations, which require periodic boundary conditions in imaginary time. DMC simulations display the same scaling both in the ferromagnetic Ising chain and in the more challenging shamrock model, as opposed to the previous finite-temperature PIMC simulations which have demonstrated to be efficient only in the former model, but are affected  in the latter by the pathological slowdown  mentioned above. Modified PIMC simulations with open boundary conditions in imaginary time for the shamrock model have not been performed yet.

The analysis of possible systematic biases of the DMC algorithm, in particular the one due to the finite random-walker population (see section ~\ref{secbias}), shows that the maximum relative error in the prediction of the ground-state energy increases with the system size. The convergence to the exact infinite random-walker number limit becomes slower as the system size increases, and the number of random walkers required to maintain a fixed relative error increases asymptotically exponentially with the system size. 
We emphasize that these findings apply to the simple DMC algorithm considered here, which represents the worst case scenario in which no suitable guiding wave function that approximates the ground state can be defined; it is possible that the importance sampling technique would boost the algorithm efficiency and fasten the convergence of the systematic biases.

The rest of the article is organized as follows: in Section~\ref{secmethod} we describe the implementation of the DMC algorithm for quantum Ising Hamiltonians. 
In Section~\ref{chain}, we describe the protocol used to measure the characteristic time of tunneling events occurring during the DMC simulations, and we provide the results for the ferromagnetic quantum Ising chain, making comparisons with exact diagonalization calculations of the gap, showing the $1/\Delta$ scaling of the tunneling time with the system size. 
In Section~\ref{secshamrock} the system-size scaling of the DMC tunneling time for the shamrock model is analyzed, showing also in this case the $1/\Delta$ scaling. 
Section~\ref{secbias} reports the analysis of the convergence of the systematic bias of the DMC algorithm due to the finite size of the random walker population.
Our conclusions and the outlook are reported in Section~\ref{secconclusions}.

\section{Diffusion Monte Carlo algorithm} \label{secmethod}
The DMC algorithm was introduced in Ref.~\cite{andersonJB}, where the analogy between the imaginary time Schr\"odinger equation and a diffusion equation was first exploited to study ground-state properties. In its many variants, it has demonstrated to be one of the most powerful computational tools to predict ground-state properties of quantum many-body Hamiltonians that describe various physical and chemical systems, including electron gases~\cite{ceperley1980ground}, electrons in atoms, molecules, and solids~\cite{reynolds1982fixed,hammond1994monte,foulkes2001quantum}, quantum fluids~\cite{boronat}, nuclear matter~\cite{carlson2003quantum}, and ultra-cold atoms~\cite{giorgini1999ground}.
In this section, we present the implementation of the DMC algorithm for transverse-field Ising models, following the theoretical formalism sketched in Ref.~\cite{nishimorifoundations}. Here, we consider a generic Ising spin Hamiltonian defined as:
\begin{equation}
\hat{H}=-\sum_{i,j} J_{ij} {\sigma}^{z}_{i} {\sigma}^{z}_{j}-\Gamma \sum_{i=1}^{N} {\sigma}^{x}_{i},
\label{H}
\end{equation}
where $J_{ij}$ is the interaction strength between the $i^{th}$ and the  $j^{th}$ spin placed on the $N$ nodes of a graph. ${\sigma}^{z}_{i}$ and ${\sigma}^{x}_{i}$ are Pauli matrices acting at site $i$. Each spin experiences a transverse field of strength $\Gamma$, which introduces quantum fluctuations. We set the reduced Planck constant $\hbar = 1$. We do not explicitly consider longitudinal fields; however, their effect could be trivially included in the algorithm.\\
The DMC algorithm projects out the ground-state wave function by evolving the time-dependent Schr\"odinger equation in imaginary time $\tau=i t$. In the Dirac notation it is given by:
\begin{eqnarray}
 -\frac{\partial}{\partial \tau}|\Psi(\tau)\rangle=(\hat{H}-E_{\mathrm{ref}})|\Psi(\tau)\rangle.
\label{DiracSHE}
\end{eqnarray}
Given $\left| s_i \right>$ an eigenstate of the Pauli matrix ${\sigma^z_i}$ at site $i$ with eigenvalue $s_i=1$ when $\left|s\right>=\left|\uparrow \right>$ and $s_i=-1$ when $\left|s \right>=\left|\downarrow \right>$, the quantum state of $N$ spins in the system is indicated by $\left|\textbf{X} \right> = \left|s_1 s_2 ... s_N\right>$. 
The ensemble of $2^N$ states $\{ |\textbf{X}\rangle \}$ is chosen as computational basis.  $\Psi({\bf X},\tau)=\langle \textbf{X}|\Psi(\tau)\rangle$ denotes the wave function at the imaginary time $\tau$. $E_{\mathrm{ref}}$ is a reference energy, which has to be adjusted to stabilize the simulation, as explained below.\\
The Schr\"odinger equation~(\ref{DiracSHE}) can be solved by applying iteratively the equation:
 \begin{equation}
 \Psi(\textbf{X},\tau+\Delta \tau)=\sum_{\textbf{X}^\prime}G(\textbf{X},\textbf{X}^\prime,\Delta \tau)\Psi(\textbf{X}^\prime,\tau),
 \label{she}
\end{equation}
where $\Delta\tau$ is a short time step, and $G({\bf X},{\bf X}^\prime,\Delta \tau)$ is the Green's function of Eq.~(\ref{DiracSHE}). In this article we employ the symmetrized primitive Trotter approximation~\cite{thijssen}:
\begin{align}
G({\bf X},{\bf X}^\prime,\Delta \tau) =    &   \\ \nonumber
 \langle \textbf{X}|e^{-\frac{\Delta \tau}{2}(\hat{H}_{\mathrm{cl}}-E_{\mathrm{ref}})}  e^{-\Delta \tau \hat{H}_{\mathrm{kin}}} & e^{-\frac{\Delta \tau}{2}(\hat{H}_{cl}-E_{\mathrm{ref}})}|\textbf{X}^\prime\rangle      +\mathrm{O}(\Delta \tau^3),
\end{align}
where $\hat{H}_{\mathrm{cl}}=-\sum_{i,j} J_{ij} {\sigma}^{z}_{i} {\sigma}^{z}_{j}$ and $\hat{H}_{\mathrm{kin}}=-\Gamma \sum_{i=1}^{N} {\sigma}^{x}_{i}$. 
By neglecting the $\mathrm{O}(\Delta \tau^3)$ terms in the Green's function, one obtains a quadratic convergence of ground-state properties in the $\Delta \tau\rightarrow 0$ limit~\cite{boronat}.  The function $G({\bf X},{\bf X}',\Delta \tau)$ is written as:
\begin{equation}
\label{primitive}
G({\bf X},{\bf X}',\Delta \tau) \approx G_d({\bf X},{\bf X}',\Delta \tau)G_b({\bf X},{\bf X}',\Delta \tau),
\end{equation}
where
\begin{equation}
G_d({\bf X},{\bf X}',\Delta \tau) = P_F^{\delta}{(1-P_F)}^{N-\delta},
 \label{diff_green}
\end{equation}
with $P_F=\frac{\sinh \left(\Delta \tau \Gamma\right) }{\exp\left({\Delta \tau \Gamma}\right)}$. $\delta$ is the number of spins with opposite orientation in ${\bf X}$ with respect to ${\bf X}'$, and
\begin{equation}
G_b({\bf X},{\bf X}',\Delta \tau)= e^{{-\Delta \tau\left[\frac{E_{cl}(\textbf{X})+E_{cl}(\textbf{X}^\prime)}{2}-N\Gamma-E_{\mathrm{ref}}\right]}}.
 \label{branching_green}
\end{equation} 
The propagator in Eq.~(\ref{diff_green}) defines a positive-definite and column-normalized (therefore stochastic) matrix. Hence, it can be used to define a conventional Markov chain. Specifically, at each iteration every spin is addressed and flipped with a probability $P_F$. Alternatively, one samples the number $\delta$ of spins to be reversed from a binomial probability distribution, and then randomly selects which spins to flip, uniformly. 
The second term $G_b({\bf X},{\bf X}',\Delta \tau)$, instead, defines a diagonal matrix which is not column normalized. Its action does not change the spin configuration. It could be accounted for by considering a large population of replicas of the system, in jargon called random walkers, and assigning to each walker a weight, which is initially equal for all walkers, and is then updated iteratively at each DMC step proportionally to $G_b({\bf X},{\bf X}',\Delta \tau)$. However, this process is known to lead to an exponentially decaying signal, since most walkers would in short imaginary time accumulate a negligible weight compared to few others. The most commonly adopted procedure to circumvent this signal loss consists in implementing a cloning/death process, in jargon called branching, in which at every iteration, say at imaginary time $\tau$, each walker generates (after the spin flips) a number of descendants for the next iteration at imaginary time  $\tau+\Delta\tau$ equal to $n_\mathrm{d} = \mathrm{int}\left(G_b({\bf X},{\bf X}',\Delta \tau)+\eta\right)$, where $\eta\in\left[0,1\right]$ is a uniform random number and the function $\mathrm{int}()$ gives the integer part of the argument. It is easily shown that on average $n_\mathrm{d}$ corresponds to the weight $G_b({\bf X},{\bf X}',\Delta \tau)$, for a sufficiently large random-walker population.\\ 
The total number of walkers does therefore fluctuate at each iteration, and after an equilibration time the walkers sample configurations according to the ground-state wave function: $\Psi({\bf X},\tau\rightarrow\infty) = \Psi_0({\bf X})$. The ground state energy and, analogously, expectation values of other operators that commute with the Hamiltonian, can be correctly estimated as 
$E=\lim_{M\rightarrow \infty} \sum_{i=1}^M E_{\mathrm{loc}}( \textbf{X}_i )/M$, 
where $\left\{\textbf{X}_i \right\}$ is a large ensemble of spin configurations generated by the DMC algorithm and  
$E_{\mathrm{loc}}(\textbf{X})=E_{\mathrm{cl}}(\textbf{X})-N\Gamma$ is the local energy. 
By tuning $E_{\mathrm{ref}}$, one can adjust the average random-walker number close to a target value $N_{\mathrm{w}}$ (in the following simply referred to as number of walkers or population size). To do so, we follow the textbook recipe described in Ref.~\cite{thijssen}. 
The correlations among different identical walkers generated in the branching process and the need to control the walker population size possibly introduce a bias, which vanishes in the $N_w \rightarrow \infty$ limit.\\

The potential sources of systematic errors in the DMC algorithm originate from the finite time step $\Delta\tau$ and the finite number of random walkers $N_{\mathrm{w}}$. 
For what concerns the DMC tunneling times, we carefully analyzed these effects, and we report in Sections \ref{chain} and \ref{secshamrock} only data obtained with small enough values of $\Delta\tau$ and large enough values of $N_{\mathrm{w}}$ to be in the asymptotic exact regime. 
For what concerns predictions of ground-state energies, a detailed analysis of the systematic bias due to the finite $N_{\mathrm{w}}$ is reported in Section~\ref{secbias}.
The systematic error in the ground-state energy due to the finite $\Delta\tau$ is less relevant and can be made smaller than statistical uncertainties with moderate computational effort, so its analysis is not reported in this article. 

Certain previous SQA studies~\cite{santoroGFMC,jordanDMC} employed an alternative projective QMC method --- which in computational condensed matter physics is usually referred to as Green's function Monte Carlo~\cite{trivedi1990ground} --- based on a different short-time approximation for the Green's function obtained with a first-order Taylor expansion: $\exp\left(-\Delta \tau ( \hat{H}-E_{\mathrm{ref}})\right) \simeq I+\Delta\tau E_{\mathrm{ref}}-\Delta\tau \hat{H}$. This method represents a stochastic implementation of the power method of linear algebra for extracting the principal eigenvector and the corresponding eigenvalue of a matrix. It converges to the exact ground state as long as the time step is small enough to ensure that the right hand side of the above Taylor expansion is always nonnegative~\cite{schmidt2005green}. This condition is easily fulfilled for small systems, but it demands smaller and smaller time-steps (and, therefore, a reduced probability to flip a spin) for larger systems, leading to very inefficient simulations (see, however, the continuous-time algorithms of Refs.~\cite{trivedi1990ground,sorella2000green}). For the same reason, this method cannot be employed in continuous-space models with unbounded spectra, as opposed to the DMC algorithm employed in this article. One relevant difference between this method and the DMC algorithm is that in the former only one spin is flipped at each iteration, while in the latter the number of spins to be flipped follows a binomial distribution, possibly making the sampling dynamics more efficient. Furthermore, in the algorithm using the Taylor expanded Green's function the variable $\tau$ no longer has the significance of imaginary time~\cite{schmidt2005green}.
%

%
%
%
\begin{figure}
\begin{center}
\includegraphics[width=1.0\columnwidth]{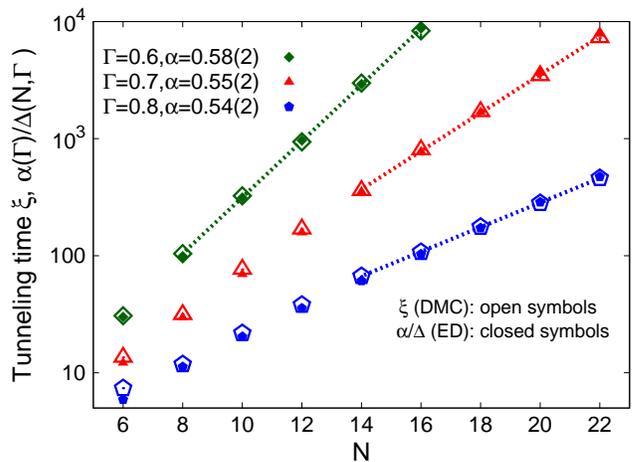}
\caption{(color online). DMC tunneling time $\xi$ for the ferromagnetic Ising chain (open symbols) as a function of the number of spins $N$, for different values of the transverse field $\Gamma$ with $J=1$. The closed symbols represent the inverse gap values $1/\Delta$ obtained with exact diagonalization and rescaled by a parameter $\alpha(\Gamma)=\mathrm{O}(1)$. The thin dashed curves represent exponential fits on the tunneling time $\xi$  in the large-$N$ regime. Here and in the other graphs the error bars are smaller than the symbol size if not visible. 
}
\label{fig1}
\end{center}
\end{figure}
%

%
%
\begin{figure}
\begin{center}
\includegraphics[width=0.6\columnwidth]{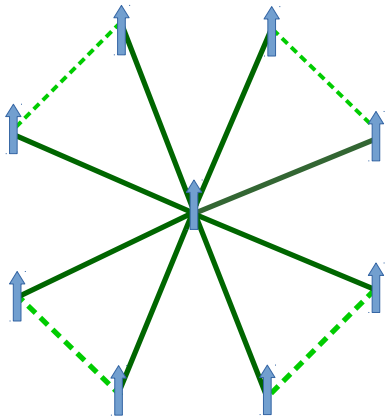}
\caption{(color online). The shamrock, a model of $N$ frustrated spins in a transverse field.  It is made up of $K=(N-1)/2$ leaves each having three spins. The solid dark-green lines depict ferromagnetic interactions (with interaction strength $J=6$) between the central spin and all the other $N-1$ spins. The dashed light-green lines instead show the anti-ferromagnetic interactions (with interaction strength $J-\epsilon$) between the outer spins of the same leaf (see Eq.~\ref{shamrock}). The overall effect results in creating $2^K$ tunneling paths between the degenerate classical ground states in the incoherent quantum tunneling regime.
}
\label{fig2}
\end{center}
\end{figure}
%

\section{The Ferromagnetic quantum Ising Chain}
\label{chain}
In this section, we describe the protocol we use to measure the characteristic time of tunneling events occurring in a DMC simulation, and we present the results for the one-dimensional ferromagnetic transverse-field Ising model defined by the following Hamiltonian:
\begin{equation}
\hat{H}=-\sum_{i=1}^N J{\sigma}^{z}_{i} {\sigma}^{z}_{i+1}-\Gamma \sum_{i=1}^{N} {\sigma}^{x}_{i},
\label{ising}
\end{equation}
where the coupling is $J > 0$ and $\Gamma$ is the intensity of the transverse field. Periodic boundary conditions are considered, i.e., ${\sigma}^{a}_{N+1}={\sigma}^{a}_{1}$ where $a=x,y,z$.\\
At zero temperature this model undergoes a quantum phase transition from a paramagnetic phase at $\Gamma > J$ to a ferromagnetic phase at $\Gamma < J$. In the $\Gamma \rightarrow 0$ limit quantum fluctuations are suppressed and one has two degenerate (classical) states with all spins up $\left|\uparrow \uparrow \dots\uparrow \right>$ or all spins down $\left|\downarrow \downarrow \dots \downarrow  \right>$. In order to go from one state to the other, the system would have to overcome an energy barrier separating the two minima, with the magnetization playing the role of a one-dimensional reaction coordinate which parametrizes a symmetric double-well profile.
For small $\Gamma > 0$, in the thermodynamic limit there are still two degenerate ground states with opposite magnetizations, but in a finite chain the degeneracy is lifted by an exponentially small (in the system size) energy gap due to the quantum tunneling which couples the two states. 
This scenario is reminiscent of what happens in a QA towards the end of the annealing process when the transverse field is small and the system explores different well-separated local minima via incoherent quantum tunneling. For this reason, shedding light on how tunneling events take place in QMC simulations --- even in the simple double-well scenario --- is important to understand if QAs have the potential to outperform classical heuristic optimization algorithms, such as SQA.\\
We define the DMC quantum tunneling time $\xi$ by implementing the following protocol: the simulation starts with all random walkers initialized in the basis state with all spins pointing up; we then measure the imaginary time $\tau$ (computed as time step $\Delta \tau$ times number of DMC iterations) required to first reach a certain percentage of walkers, somewhat arbitrarily taken to be $25\%$, with negative magnetization (majority of spins pointing down), meaning that they have crossed the energy barrier. This definition is analogous to the one employed in Refs.~\cite{isakovtunneling,mazzolaquantumchemistry,aminshamrock} in the case of PIMC simulations, where a certain percentage of imaginary-time slices, instead of walkers, is considered. The simulation is repeated approximately $250$ times for larger systems and small $\Gamma$ and approximately $2500$ for smaller systems and larger values of $\Gamma$. We then take the average value to define $\xi$ and its standard deviation to define the error bar.\\
The DMC tunneling times for the ferromagnetic Ising chain are shown in Fig.~\ref{fig1}, as a function of the number of spins $N$ and for different values of $\Gamma$. For large $N$ the data display an exponential growth, quite similar to the dependence of the inverse gap $1/\Delta$, which we obtain via exact diagonalization of the Hamiltonian matrix. In fact, by multiplying the inverse gap $1/\Delta$ by an appropriate numerical prefactor $\alpha$ we obtain precise matching between the two datasets. The coefficient $\alpha$ turns out to be a number $\mathrm{O}(1)$. We also consider different definitions of DMC tunneling time, using percentages of walkers that have to cross the barrier between $10\%$ and $25\%$, obtaining again results which follow the $1/\Delta$ scaling but with a slightly different value of the prefactor $\alpha$.\\
The inverse-gap scaling displayed by the DMC tunneling times is similar to the result found in Ref.~\cite{isakovtunneling} using modified PIMC simulations performed using open boundary conditions in imaginary time. This is not surprising, since such modified PIMC method had been originally introduced as a computational tool to study ground-state properties~\cite{sarsa,boninsegniPigs}. However, it is usually employed in combination with guiding wave functions that accurately describe the ground state, so that the convergence to the zero-temperature limit as a function of the total path length is quite rapid. How this algorithm converges to the ground state in the absence of the guiding wave function has not been analyzed in detail yet.
It is also worth stressing that in the PIMC formalism the tunneling time is defined by counting the number of Monte Carlo sweeps (a sweep corresponds to one Monte Carlo step per spin) and, therefore, it does not bear the significance of imaginary time as in the DMC method employed in this article.
In Ref.~\cite{isakovtunneling}, also finite-temperature PIMC simulations (with periodic boundary conditions) have been performed, finding that the PIMC tunneling times scale as $1/\Delta^2$. This behavior was found in ferromagnetic Ising models, which are characterized by a one-dimensional reaction coordinate, and it was later confirmed also  in one-dimensional and two-dimensional continuous-space models~\cite{mazzolaquantumchemistry}, showing that it persists also when the reaction coordinate in multidimensional.\\
Considered together, the above findings suggest that QMC algorithms are either as efficient as (in the case of the finite-temperature PIMC algorithm) or quadratically faster than (in the case of the PIMC algorithm with open boundary conditions in imaginary time or of the DMC algorithm) QAs in tunneling through energy barriers and therefore, if one assumes that incoherent quantum tunneling is the major resource of QAs, also in solving optimization problems.
%

%

%
%
\begin{figure}
\begin{center}
\includegraphics[width=1.0\columnwidth]{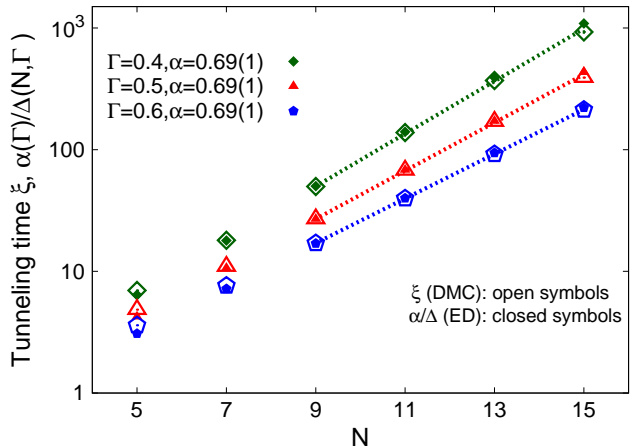}
\caption{(color online). DMC tunneling time $\xi$ for the shamrock model DMC (open symbols) as a function of the number of spins $N$, for different values of the transverse field $\Gamma$.  The other system parameters are $J=6$ and $\epsilon=0.2$. The filled symbols represent the inverse gap values $1/\Delta$ obtained with exact diagonalization and rescaled by a parameter $\alpha(\Gamma)=\mathrm{O}(1)$. The thin dashed curves are exponential fits to the tunneling time $\xi$ in the large-$N$ regime.
}
\label{fig3}
\end{center}
\end{figure}
%

%
\begin{figure}
\begin{center}
\includegraphics[width=1.0\columnwidth]{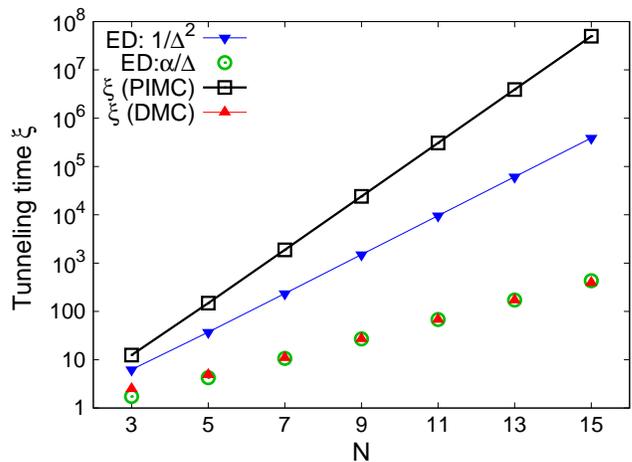}
\caption{(color online). Comparison between the tunneling times of the finite-temperature PIMC algorithm and the DMC algorithm, for the shamrock model at  $\Gamma=0.5$. 
The PIMC data are obtained from the formula $\xi=2^K/\Delta^2$, which was found in Ref~\cite{aminshamrock}. The solid blue points represent the scaling $1/\Delta^2$, characteristic of incoherent quantum tunneling.
The red triangles represent the DMC data. They are well described by the scaling law $\alpha/\Delta$ (green empty circles), where the gap $\Delta$ is obtained via exact diagonalization.
The simulation parameters are $J=6$ and $\epsilon=0.2$.
}
\label{fig4}
\end{center}
\end{figure}

\section{The shamrock model}
\label{secshamrock}
The results for the ferromagnetic Ising chain presented in the previous section indicate that, in an effective double-well system, QMC simulations can efficiently simulate incoherent quantum tunneling and, therefore, they might potentially be as efficient as, or even faster than QAs in solving complex optimization problems. In order to understand if this finding is valid in a more general setup, the authors of Ref.~\cite{aminshamrock} considered a model, named ``shamrock'', which contains the minimal elements of frustration. This model is described by the following Hamiltonian:
 \begin{equation}
\hat{H}=-J\sum_{i=1}^{K}\sum_{j=2i}^{2i+1} {\sigma}^{z}_{1} {\sigma}^{z}_{j}+(J-\epsilon)\sum_{i=1}^{K}  {\sigma}^{z}_{2i} {\sigma}^{z}_{2i+1}  -\Gamma \sum_{i=1}^{N} {\sigma}^{x}_{i}.
\label{shamrock}
\end{equation}
The $N$ spins are grouped in $K$ rings, which form the leaves of the shamrock. See Fig.~\ref{fig2}. Each ring is made of three spins and the $K$ rings all share one spin, which is placed in the center. The number of spins is related to the number of rings by the formula $N=2K+1$. In Eq.~(\ref{shamrock})  $\epsilon\ll J$ is a small interaction energy. The first term in Eq.~(\ref{shamrock}) describes ferromagnetic interactions between the central spin and the outer $N-1$ spins. The two outer spins of each ring are coupled to each other by an anti-ferromagnetic interaction, described by the second term in Eq.~(\ref{shamrock}). The intensity of the transverse field in the last term in Eq.~(\ref{shamrock}) is $\Gamma$.\\

We investigate the DMC tunneling time using the protocol described in Section~\ref{chain}. The results are shown in Fig.~\ref{fig3}. They display the same $1/\Delta$ scaling already observed in the case of the ferromagnetic Ising chain, corresponding to a quadratic speedup with respect to incoherent quantum tunneling. The value of the prefactor $\alpha$ used to superimpose the inverse-gap data to the DMC tunneling time is, as in the ferromagnetic Ising chain, a number of $\mathrm{O}(1)$. This suggests that frustrated couplings do not play a fundamental role in the tunneling dynamics of DMC simulations.\\
In Fig.~\ref{fig4} we also report the scaling of the tunneling times $\xi$ obtained in finite-temperature PIMC simulations in Ref.~\cite{aminshamrock}. As opposed to the DMC data, which display the same $1/\Delta$ scaling in the ferromagnetic Ising chain and in the shamrock model, the PIMC results display, in the latter model, a faster growth of $\xi$ with the system size, very accurately described by the scaling law $\xi^{PIMC} \propto 2^K/{\Delta^{2}}$. Due to the $2^K$ term, this growth is exponentially faster than the scaling of the DMC tunneling time and of the timescale of incoherent quantum tunneling. This pathological slowdown of PIMC simulations was indeed anticipated by the perturbation theory of Ref.~\cite{aminshamrock}. This theory predicts that in frustrated models where the two competing ground states are connected by a number of homotopy-inequivalent paths which grows with system size, incoherent quantum tunneling can display a quantum speedup if many inter-path transitions are inhibited by topological obstructions (related to the obstructions discussed in Ref.~\cite{hastings2013}). The shamrock model was indeed introduced as an example of this scenario, with the PIMC simulations confirming the theoretical prediction also beyond the perturbative regime.

\begin{figure}
\begin{center}
\includegraphics[width=1.0\columnwidth]{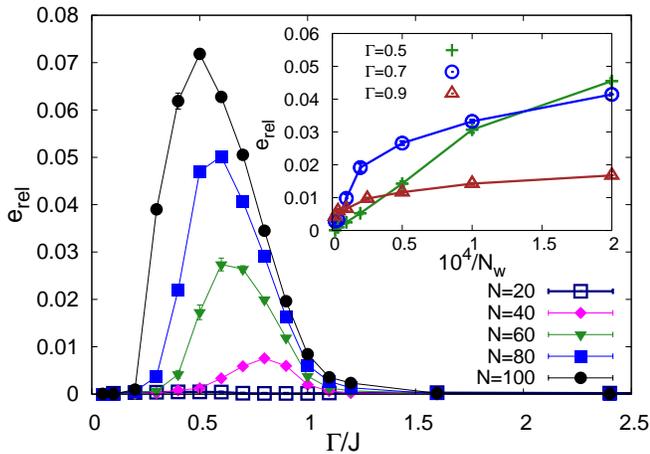}
\caption{(color online). Main panel: Relative error $ e_{\mathrm{rel}} = \left|E - E_{\mathrm{JW}}\right| / \left|E_{\mathrm{JW}}\right|$ of the DMC result $E$ with respect to the exact Jordan--Wigner theory $E_{\mathrm{JW}}$ as a function of the transverse field intensity $\Gamma$, for different system sizes $N$. The average number of random walkers is $N_w=20000$. 
Inset: $ e_{\mathrm{rel}}$ as a function of the rescaled inverse number of walkers $1/N_w$, for different transverse field intensity $\Gamma$. The size of the spin chain is N=60.
}
\label{fig5}
\end{center}
\end{figure}

\begin{figure}
\begin{center}
\includegraphics[width=1.0\columnwidth]{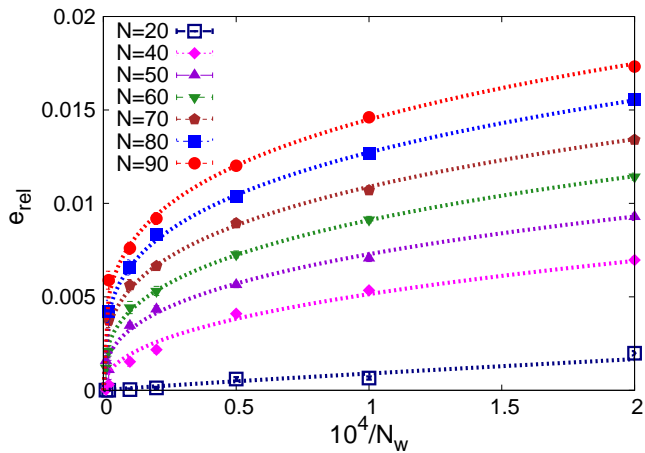}
\caption{(color online). 
Relative error $ e_{\mathrm{rel}}$ as a function of the rescaled inverse number of walkers $1/N_w$, for different system sizes. The transverse field intensity is $\Gamma = 0.95J$. The dashed curves represent power-law fitting functions (see text).
}
\label{fig6}
\end{center}
\end{figure}

\section{Analysis of the systematic bias in DMC simulations due to the finite random-walker population}
\label{secbias}
\begin{figure}
\begin{center}
\includegraphics[width=1.0\columnwidth]{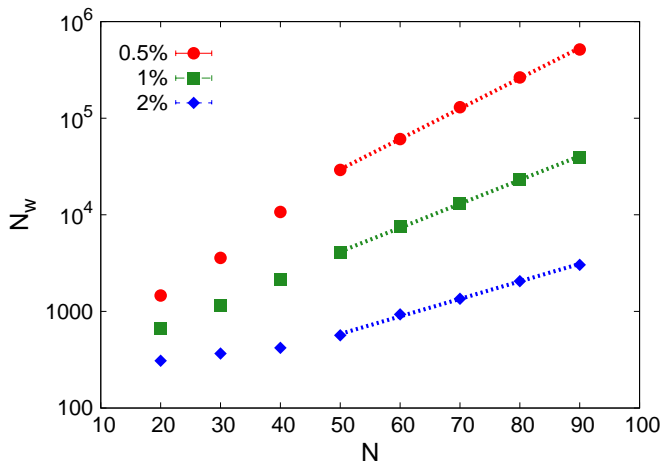}
\caption{(color online). 
Random-walker number necessary to have $0.5\%$, $1\%$, and $2\%$ relative error as a function of the system size $N$. The transverse field intensity is $\Gamma = 0.95J$.
}
\label{fig7}
\end{center}
\end{figure}

As explained in Section~\ref{secmethod}, the ground-state energy obtained via DMC simulations is subject to two sources of possible systematic bias, originating from the finite time step $\Delta \tau$ and from the finite random-walker number $N_w$. The convergence to the $\Delta \tau$ limit is quadratic, and all results presented in this paper have been performed using sufficiently small $\Delta \tau$ to make its systematic effect negligible compared to the statistical uncertainty. It this section, we focus on the bias resulting from the finite value of $N_w$.\\
We consider the ferromagnetic quantum Ising chain defined in Eq.~(\ref{ising}). Its ground-state energy per site can be exactly determined via Jordan--Wigner transformation, obtaining $E_{\mathrm{JW}}/N = -\frac{2J}{\pi}\left(1+\Gamma/J\right)\mathrm{E}\left(\pi/2,\theta\right)$, where $\mathrm{E}(x)$ is the complete elliptic integral of the second kind and $\theta^2=4\Gamma/\left[J\left(1+\Gamma/J\right)^2\right]$~\cite{pfeuty1970one}. This results correspond to the thermodynamic limit. The results presented in this section have been obtained using sufficiently large system sizes so that finite-size effects are not relevant.\\
In the main panel of Fig.~\ref{fig5}, we plot the relative error $ e_{\mathrm{rel}} = \left|E - E_{\mathrm{JW}}\right| / \left|E_{\mathrm{JW}}\right|$ of the DMC result $E$ with respect to the Jordan--Wigner theory as a function of the transverse field intensity, for different system sizes. These data correspond to a fixed random walker population $N_w=20000$. One notices that in the paramagnetic phase $\Gamma > J$, as well as in the $\Gamma \rightarrow 0$ limit, the systematic bias due to the finite $N_w$ is negligible. However, in the ferromagnetic phase $0 < \Gamma < J$ a systematic bias is observable, and this bias increases with the system size $N$. 
In order to remove a potential doubt as of why the maximum value of $e_{\mathrm{rel}}$ seems to be shifting towards lower values of $\Gamma$, we have included an inset in Fig.~\ref{fig5}, showing $e_{\mathrm{rel}}$ with respect to a rescaled $1/N_w$ for a chain of $60$ spins. As expected, in the infinite $N_w$ limit, the hardest point in the phase diagram to simulate is at the left of the quantum critical point. This should probably be due to the fact that in the vicinity of the phase transition, the large quantum fluctuations require very large $N_w$ of the DMC algorithm without importance sampling in order to capture in a reasonable accuracy the ground state properties of the system.\\
To better understand the effect of the finite random-walker population, we analyze in Fig.~\ref{fig6} the convergence to the exact Jordan--Wigner result in the $N_w\rightarrow\infty$ limit, considering different system sizes, at $\Gamma=0.95$. The data are well described by power-law fitting functions of the type $e_{rel}= c/N_w^{\beta}$, where $c$ and ${\beta}$ are fitting parameters. 
The exponent ${\beta}$ decreases with the systems size, meaning that, as the system size increases, it takes a larger population of walkers to obtain accurate predictions. In order to quantify this dependence, in Fig.~\ref{fig7} we show how the number of walkers required to have a fixed relative error increases with the system size. In the large-$N$ limit, the data are well described by an exponential fitting function, possibly indicating that the computational complexity of the simple DMC algorithm (i.e., without the use of the importance sampling technique) is exponential in the system size.

%
%
\section{Conclusions and outlook}
\label{secconclusions}
We implemented a projective QMC method for quantum Ising models based on the DMC algorithm --- in which the transition matrix is defined using a Trotter approximation of the Green's function --- and we investigated the characteristic time of tunneling events in problems characterized by an effective double-well energy landscape. We found that the DMC tunneling time increases with the system size as the inverse of the gap, that is, more favorably than the incoherent tunneling time, which increases as the inverse gap squared. This inverse-gap scaling was found to hold both for a ferromagnetic quantum Ising chain and for the more challenging shamrock model. This is in contrast with previous studies based on finite-temperature PIMC simulations, where a pathological slowdown due to topological obstructions originating from frustrated interactions was found to cause, in the shamrock model, an increase of the PIMC tunneling time which is exponentially faster~\cite{aminshamrock} than the inverse-gap squared scaling observed in the case of simple ferromagnetic models~\cite{isakovtunneling}.  
%
Our findings indicate that the DMC algorithm is not affected by the obstructions that slow down the PIMC tunneling dynamics,  thus suggesting that this algorithm is a more efficient engine for SQA considered as a heuristic optimization method.\\

Motivated by the arguments of Ref.~\cite{aminshamrock} --- according to which a classical algorithm is to be considered an efficient simulation of QAs only if it reproduces both their tunneling dynamics and their equilibrium properties --- we analyzed the computation time required by the DMC algorithm to accurately predict ground-state properties.
The analysis of the systematic bias in the ground-state energy due to the finite random-walker population revealed an exponential increase of the population size and, therefore, of the computation time, required to keep a fixed relative error as the system size increases. 
This suggests that, in general, the computational effort required to simulate the behavior of QAs via simple DMC simulations running on classical computers scales exponentially, leaving the door open for potential quantum speedup.\\
The finding of this exponential scaling is consistent with the statement of Ref.~\cite{bravyi2015monte} that the problem of estimating the ground state energy of a stoquastic Hamiltonian with a small additive error is at least $\mathrm{NP}$-hard. This statement is based on the observation that any Hamiltonian diagonal in the computational basis is stoquastic, and that finding its ground state encompasses hard optimization problems such as $k$-SAT and MAX-CUT. This essentially rules out the possibility that a polynomially scaling algorithm applicable to  generic stoquastic Hamiltonians can be found.
Still, for certain ferromagnetic models, including the transverse-field Ising chain considered in this article, algorithms which --- albeit being far from practical --- have a provably polynomial scaling have recently been found~\cite{bravyi2017polynomial}. However, since the DMC algorithm we employ in this article is not tailored to a specific (e.g., ferromagnetic) Ising model, it is natural to observe the exponential behavior corresponding to a generic model.\\

We stress once again that the above-mentioned findings correspond to the simple DMC algorithm considered in this article. It is plausible that the computational cost could be drastically reduced by using importance sampling techniques based on suitably constructed guiding wave functions, possibly at the point of modifying the scaling of the required random-walker population, at least in cases where accurate approximations of the ground-state wave functions can be constructed.
The use of importance sampling might also allow one to efficiently simulate the models described in Refs.~\cite{jordanDMC,bringewatt2017diffusion}, for which simple (i.e., without importance sampling) projective QMC methods have been shown to fail due to the large discrepancy between the $L_1$-normalized ground-state wave function --- which is the probability distribution sampled from in simple projective QMC simulations --- and the $L_2$-normalized ground-state wave function, which is sampled from when performing a measurement on the ground state of the adiabatic process.
We plan to investigate these issues in future works.\\
We argue that finding models where such importance sampling technique is not feasible (e.g., because no accurate and efficiently computable guiding wave function exists) could help us in identifying optimization problems where quantum advantage can be achieved. For the same purpose, it would be useful to identify which features of a Hamiltonian might cause a pathological slowdown of the DMC dynamics.

We acknowledge insightful discussions with Guglielmo Mazzola and Rosario Fazio. S. P. acknowledges financial support from the BIRD2016 project ``Superfluid properties of Fermi gases in optical potentials'' of the University of Padova.


%

\end{document}